\documentclass[a4paper,10pt]{article}
\usepackage[utf8x]{inputenc}
\usepackage{graphicx,amsmath,amsfonts}
\usepackage{authblk}

\title{A physical theory of economic growth}

\author{Hans G. Danielmeyer}
\author{Thomas Martinetz}
\affil{Institut f\"ur Neuro- und Bioinformatik, Universit\"at zu L\"ubeck, Germany}

\begin{document}

\maketitle

\begin{abstract}
Economic growth is unpredictable unless demand is quantified. We solve this problem by introducing the demand for unpaid spare time and a user quantity named human capacity. It organizes and amplifies spare time required for enjoying affluence like physical capital, the technical infrastructure for production, organizes and amplifies working time for supply. The sum of annual spare and working time is fixed by the universal flow of time. This yields the first macroeconomic equilibrium condition. Both storable quantities form stabilizing feedback loops. They are driven with the general and technical knowledge embodied with parts of the supply by education and construction. Linear amplification yields S-functions as only analytic solutions. Destructible physical capital controls medium-term recoveries from disaster. Indestructible human capacity controls the collective long-term industrial evolution. It is immune even to world wars and runs from 1800 to date parallel to the unisex life expectancy in the pioneering nations. This is the first quantitative information on long-term demand. The theory is self-consistent. It reproduces all peaceful data from 1800 to date without adjustable parameter. It has full forecasting power since the decisive parameters are constants of the human species. They predict an asymptotic maximum for the economic level per capita. Long-term economic growth appears as a part of natural science.
\end{abstract}

\section{Introduction}
Detailed production values are obtained at the microeconomic level from precise bookkeeping required for taxation. They are integrated to the macroeconomic national level, carefully corrected for inflation, reduced to per capita values, and published in the national statistical reports. We use them of course. 

Unfortunately there is no such method for quantifying the user side. Its values are personal, partially emotional, and cannot be added to the national level because they overlap in content and user time. This absolute problem of economics explains why the equilibrium between demand and supply could still not be quantified. There exists not even a compatible set of variables for both sides. These deficits allowed the growth fetishism and uncontrolled monetarism that caused the banking disaster of 2008 and the largest gap between public debt and private fortunes of all times. 

The other fundamental problems of macroeconomic theory are due to the fact that its conceptual frame was also derived from microeconomics. It began in 1776 with Adam Smith’s invention of an “invisible hand” for achieving macroeconomic equilibrium [1], and in 1803 with Jean-Baptiste Say’s unidirectional claim that every product creates its market [2]. 

The most elaborate frame was based on Robert M. Solow’s neoclassical exponential growth theory of 1956 [3]. Robert J. Barro and Xavier Sala-I-Martin presented a comprehensive quantitative analysis of neoclassical production functions and medium-term growth patterns [4]. It leads directly to the central issue for our physical theory. The neoclassical approximation uses the mathematical tool of partial derivatives with fractional exponents for optimizing the input variables labor and physical capital (PC), the value of the technical infrastructure. But the condition sine qua non for using partial derivatives is the independence of the input variables, whereas successful nations can stabilize employment at half the population or about one person per household. This means macroeconomic employment is not even a variable, and the national value of paid labor cannot be independent because the people must be able to buy the annual output called gross domestic product (GDP).

Labor is the central quantity for any theory because in the end all economic values consist of labor, including PC and natural resources. Annual working time is the last choice and a good variable for labor since it is nationally normalized across business sectors. But industrial experience shows that the employee’s division of annual working time is designed into every piece of physical capital from ploughs to hospitals according to its operating conditions and the technical state of the art.
  
This hidden relation between annual working time and physical capital is not only the key to a physically satisfactory theory of economic growth. Simultaneously it will resolve the absolute problem of economics because in the end working time and GDP must depend on the user’s demand and the producer’s PC. But the relation’s shape could not even be guessed because PC and GDP were too often reset by the Civil War of the USA, both World Wars, the Great Depression, oil price shocks, and recently by the financial disaster of 2008.

The next section shows that these disasters left hardly a trace in the long-term decrease of the working time from 18th century to date. PC has an effective lifetime against obsolescence and decay of about 25 years so that it needs more than one reproduction cycle for recovery to long-term equilibrium. The GDP will at first be dominated by this recovery, but thereafter obviously by a second storable quantity that controls the industrial evolution above all national disasters. Identifying this quantity is the central issue for the physical theory.

Discovering a hidden relation in a system with two coexisting dynamics but missing variables, parameters, and data for one of them seems difficult. But a standard method of physics postulates the existence of missing quantities, quantifies their properties and interactions, and confirms them with the agreement between all available data and the entire theoretical frame.

The next section introduces the system’s complete set of physical variables, sets up a continuously valid equilibrium condition between them, and specifies their initial and final states. Section 3 quantifies the intrinsic feedback loops for both storable quantities. Section 4 discusses the results. 

\section{Physical variables, the universal trade-off, and economic equilibrium}

\begin{figure}
\centering  
\includegraphics[width=100 mm]{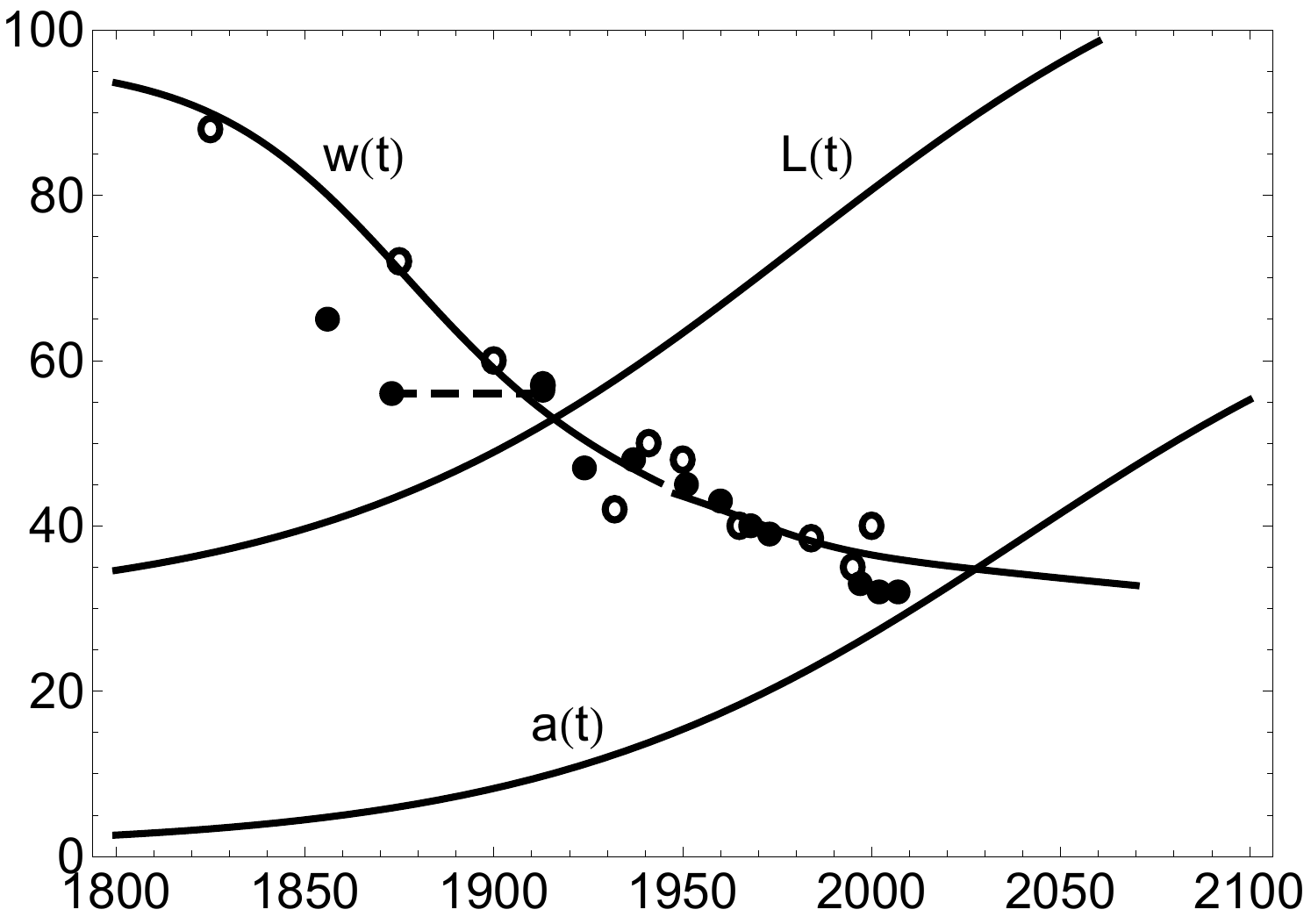}
\caption[The cultural variables]{\label{fig_1} The physical theory's cultural variables derived and plotted without any adjustable parameter. The official data for the annual working time $w(t)$ in the UK (points) and Germany (circles) are in hours per week. The data for the life expectancy $L(t)$ in years and the industrial evolution $a(t)$ in 1000 US\$ of 1991 p. a. per capita are shown in figure \ref{fig_4}.}
\end{figure}

Figure \ref{fig_1} shows the official annual working times $w(t)$ for the UK and Germany from 18th century to date as examples of the first and the most unstable industrial nation, respectively. Even World War II left hardly a trace. This confirms the industrial experience that $w(t)$ represents the state of the art of the industrial society's leading nations. It is anti-correlated with the other cultural variables. They are shown with their data in figure \ref{fig_4}. 

The maximum available working time per employee is 16 hours per day for 6 days per week or 96 hours per week. This was according to figure \ref{fig_1} also the official working time around about 1800. Since all macroeconomic changes are per year this annual maximum is the natural unit for measuring working time:
\begin{equation}
\label{eq_1}
\overline{\varepsilon}=1\; p.a. = 96 \text{ hours of work per week.}         
\end{equation}                                    
The user side’s natural counterpart of annual working time is annual spare time $s(t)$ at home. In economic theory it doesn’t turn up because it has no economic value, but G7 affluence would be unbearable without the spare time for enjoying it. Physically it is decisive since there is only one and the same time passing by for the sum of spare and working time: 
\begin{equation}
\label{eq_2}
s+w=\overline{\varepsilon}.         
\end{equation} 
Either a nation produces more or has more time to enjoy whatever it produces. In hindsight it is surprising that it took so long for replacing Say’s unidirectional theorem with a universal trade-off between both sides of the economy.

The division of paid working time and unpaid homework is within most households on average socially balanced. The second long-term fact is also observed: Decreasing working time means for a nearly constant employment level that an increasing part of PC is automated or operated only parttime. The GDP $y(t)$ is, therefore, proportional to annual working time and the part $k_w(t)$ of PC that is employed in production: 
\begin{equation}
 \label{eq_3}
y=wk_w.         
\end{equation}   
This departure from the general rule of not splitting macroeconomic quantities is required for the total value $k(t)$ of PC because the post World War II generation produced an increasing part $k_s(t)$ of PC for better living conditions at home. In the UK and Germany this part contributes now about 40 \% to the accumulated PC but after the year of construction  little to the annual GDP. 

(\ref{eq_3}) is not yet the required production function because $w(t)$ cannot be independent when it designed into $k_w(t)$. On the other hand there is no reason for assuming any non-linearity because there are no diminishing returns for supervised or operated $k_w(t)$. Finally the unit chosen with (\ref{eq_1}) for measuring working time does not require an adjustable parameter since $n=\overline{\varepsilon}/w=\overline{\varepsilon} k_w/y$ corresponds to the number of years it would take to produce exclusively $k_w(t)$ with the same productivity of producing the GDP.

The hidden relation between $w(t)$ and other variables must be controlled by a second storable quantity that organizes and amplifies spare time for demand like $k_w(t)$ organizes and amplifies $w(t)$ for supply. This new quantity is the natural counterpart of PC. It was named “human capacity” $h(t)$ [5]. 

Like PC embodies the industrial society’s technical knowledge by design, human capacity (HC) embodies the general knowledge distributed in every individual by education.  HC includes also the heritable capacities like curiosity, talent, and limited adaptability. They have no direct economic value since they are provided without extra cost, but they are indestructible and physically as relevant as spare time. The educated values of HC per capita are also indestructible when education continues in times of war. 

In analogy to (\ref{eq_3}) the user side is quantified with
\begin{equation}
 \label{eq_4}
sh_s=y         
\end{equation}                                                            
where $h_s(t)$ is that part of HC that is actually needed for enjoying but possibly also limiting the useful goods and services of the GDP. Improving and maintaining housing and working conditions, public order and social security are as important for the existential conditions as classical consumption.

Combining (\ref{eq_2}), (\ref{eq_3}), and (\ref{eq_4}) yields the industrial society’s equilibrium condition 
\begin{equation}
 \label{eq_5}
(\overline{\varepsilon}-w)h_s=y=wk_w.         
\end{equation}        
The trade-off (\ref{eq_2}) between spare and working time causes $w(t)$ to appear also on the user side. This is the hidden relation. It connects demand with the required work for economic equilibrium, but also supply with the required spare time. 

That this dual physical symmetry cannot be expressed with production variables alone explains every problem encountered to date with macroeconomic theory. But a second asymmetry remains: PC can be duplicated fast just with money and work whereas HC can only be increased slowly because it requires research for new relevant knowledge and time for human adjustment. 

It is a simple exercise to generalize (\ref{eq_5}) for including the initial state 
\begin{equation}
 \label{eq_6}
h_o=y_o/\overline{\varepsilon},\; y_o,\; k_o=y_o/\overline{\varepsilon},\; w_o=\overline{\varepsilon},\; s_o=0         
\end{equation}                                            
as minimum condition for successful reproduction. Then the generalized form of (\ref{eq_5}) reads $(\overline{\varepsilon}-w)h_s+y_ow/\overline{\varepsilon}=y=wk_w$. This includes the transition from medieval agriculture to the industrial society. Resolving it for the dependent quantities yields 
\begin{equation}
 \label{eq_7}
w=\frac{\overline{\varepsilon}h_s}{h_s + k_w - y_o/\overline{\varepsilon}}
\end{equation}                                                    
and  
\begin{equation}
 \label{eq_8}
y=\frac{\overline{\varepsilon}h_sk_w}{h_s + k_w - y_o/\overline{\varepsilon}}.
\end{equation}                                                     
Since they are continuously measued the storable quantities follow from resolving (\ref{eq_7}) and (\ref{eq_8}) for  $h_s$ and $k_w$.
      
$w(t)$ is plotted in figure \ref{fig_1} from (\ref{eq_3}) with $y_o=900$ US\$ p.a. in US\$ of 1991 and with the next section's results for PC and the GDP. After World War II it agrees with $k_w=k-k_s$ obtained from the reported data for $k_s$ and $k$. 
 
(\ref{eq_8}) is a valid production function and without adjustable parameter physically correct. HC and PC must increase together for increasing the GDP. The neoclassical phenomenon of diminishing returns does not exist since the amplification of both inputs remains linear. Spare time, $k_s$, and $y(t)$ are just the winners of more PC and HC until HC limits the GDP according to 
\begin{equation}
 \label{eq_9}
y=\frac{\overline{\varepsilon}h_s}{1 + h_s/k_w} <\overline{\varepsilon}h_s<\overline{\varepsilon}h.
\end{equation}
The next section's comparison with educational data yields $h_s(t)\approx (1/3)h(t)$. An upper limit to HC per capita is physically inevitable since the human species is not omnipotent. This has nothing to do with money, global resources, or understanding and using the laws of nature. The only question is when and how the economy must yield to mankind's generic limit. It was raised and qualitatively answered in 1930 by John Maynard Keynes [6]. The next section gives the first quantitative answer.  

\section{Physical constants, stabilizing feedback loops, analytic solutions, and economic growth}

The preceding section quantified the generation of the GDP by amplification of spare and working time with actually required parts of HC and PC. This section quantifies the opposite processes, the generation of HC and PC from the GDP with parts of annual working time. This forms a pair of closed feedback loops. They stabilize peaceful medium and long-term economic growth. 

Detailed bookkeeping uses the fiscal depreciation rates of PC. They allow tax-free maintenance of PC and stimulate economic growth because the time for total depreciation is on purpose shorter than the physical lifetime of PC. For instance, Japan allowed writing off the full value of the PC used for research and development within the first year. Economic theory uses the physical decay rate of PC because it means money. Physical theory must use the inverse of the decay rate, the physical lifetime of PC against decay and obsolescence. Storage times cause memory and different time dependencies between the stored content and its charging and discharging flows. This excludes the exponential function as legitimate growth law: It is the only function that can neither cause nor show these differences because it features unlimited exchangeability between amplitude and time. 

The physical lifetime of HC 
 \begin{equation}
 \label{eq_10}
E=62\text{ years}
\end{equation}   
was named evolution constant, the physical lifetime of PC 
 \begin{equation}
 \label{eq_11}
G=25\text{ years}
\end{equation}
generation constant [5]. They are systemic constants of the industrial society and are directly measured with the national time shifts between GDP and PC according to the following theory and their agreement with the data.

These lifetimes yield the parts of the GDP required for maintaining the values of the storable quantities with the maintenance conditions 
\begin{equation}
 \label{eq_12}
h/E=\nu y
\end{equation}
and 
\begin{equation}
 \label{eq_13}
k/G=\mu y
\end{equation}
They define the “capacity function” $\nu(t)$ and the “capital function” $\mu(t)$. 

\begin{figure}
\centering 
\includegraphics[width=100 mm]{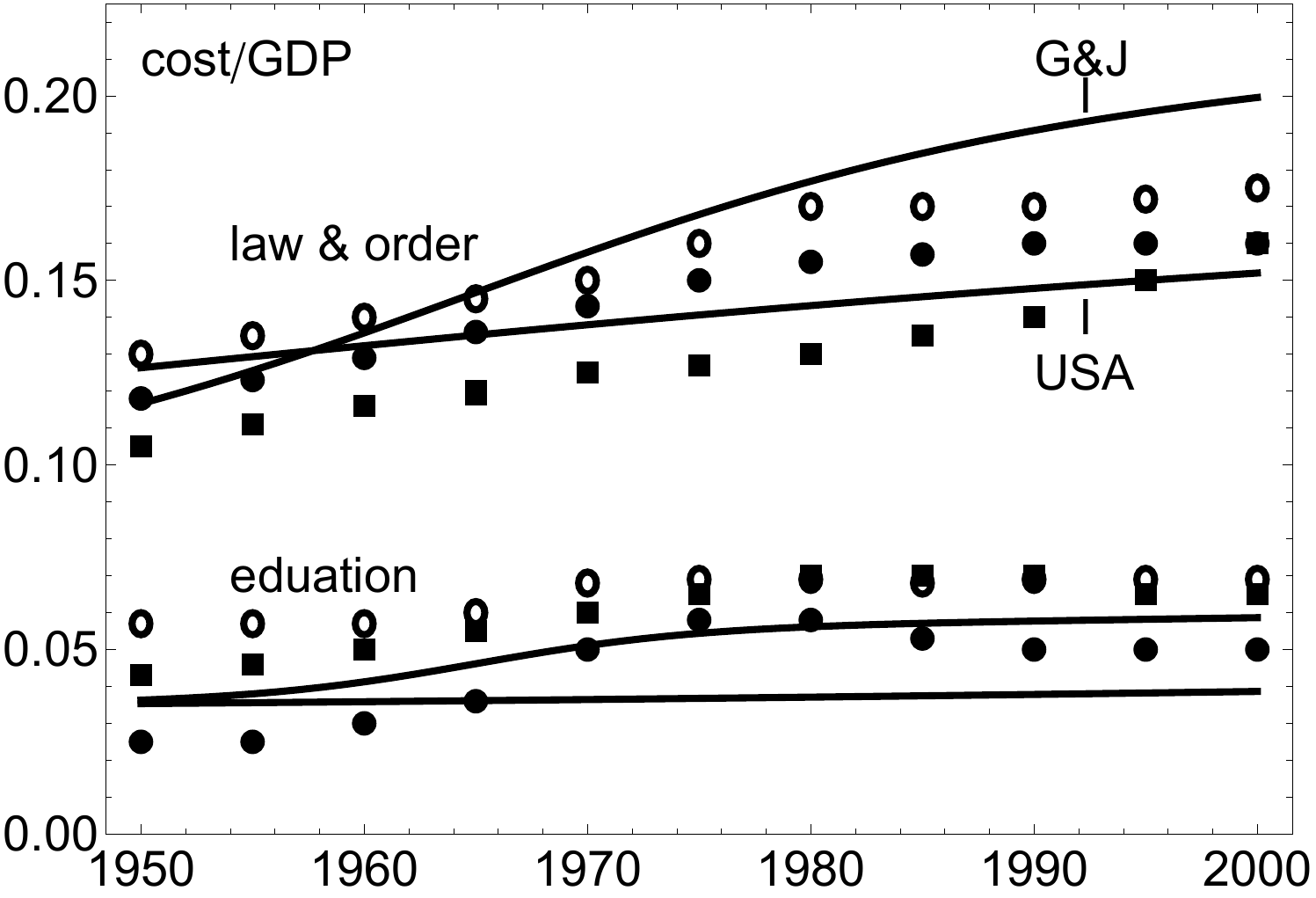}
\caption[The operational parts of the GDP]{\label{fig_2} The industrial society’s operational support. The data for education and public order for the USA (circles), Japan (squares) and Germany (points) are compared with the  capital function for Germany, Japan, and the USA (top), and  with the capacity function for the G7 (bottom). Its increase after 1958 was the G7 answer to the Soviet Union’s orbiting of Sputnik, the first satellite.}
\end{figure}

\begin{figure}
\centering  
\includegraphics[width=100 mm]{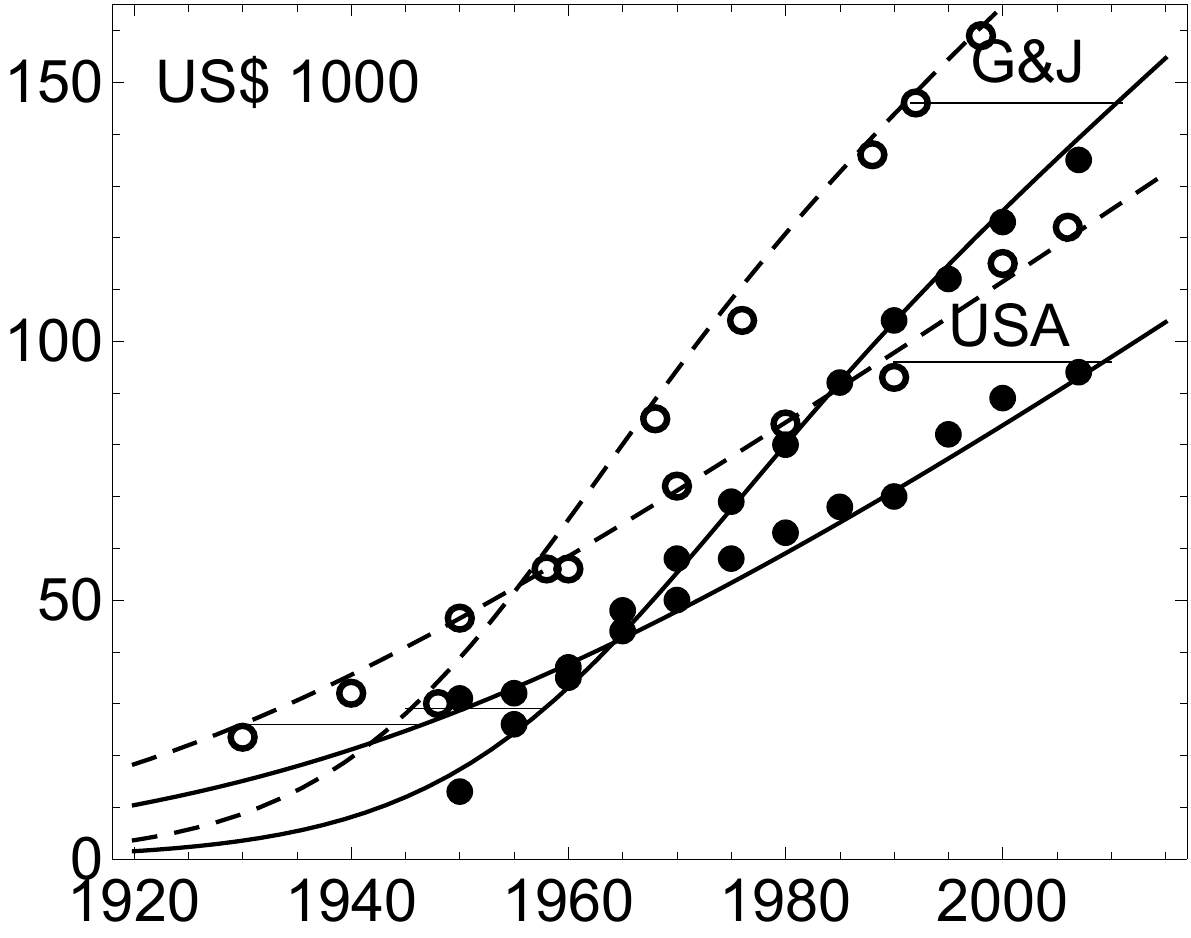}
\caption[Time shifts between GDP and physical capital]{\label{fig_3} Measuring the generation constant $G$ and the evolution constant $E$. They fix the time shifts between physical capital (points in 1000 US\$ of 1991) and the GDP normalized to the same amplitude (circles). For the industrial evolution the shift is always 21 years (top). For recovery the shift depends also on the nation's support for physical capital (bottom).}
\end{figure}

\begin{figure}
\centering 
\includegraphics[width=100 mm]{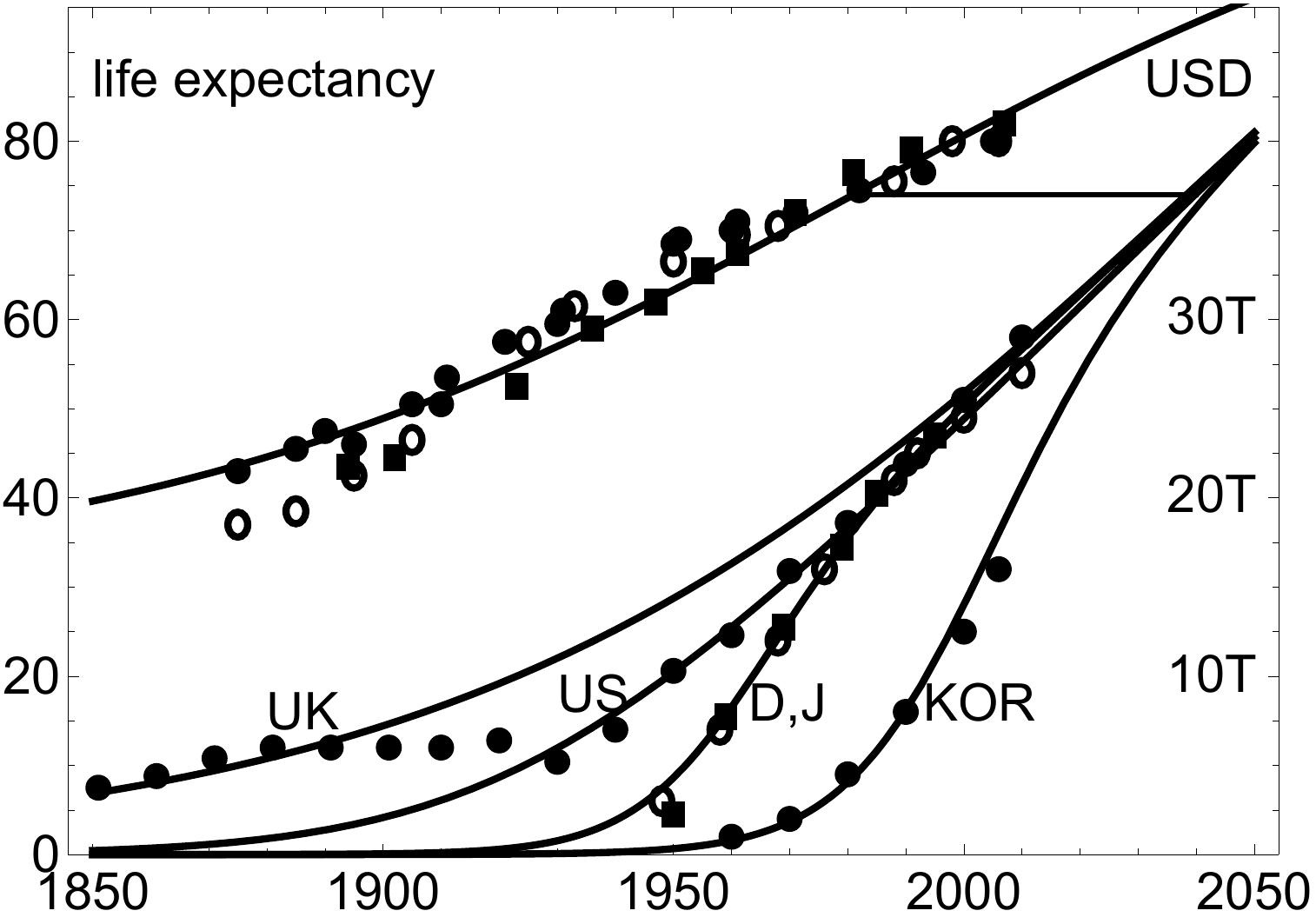}
\caption[Long-term growth and the life expectancy]{\label{fig_4} National recoveries (right hand scale), their convergence with the industrial evolution as peaceful envelope above all disasters, and the unisex life expectancy (left hand scale) for the USA (points), Germany (circles), and Japan (squares). Korea is presently at its halftime of convergence. The bar indicates the time shift of 59 years between the half times of the life expectancy and the industrial evolution.}
\end{figure}

Figure \ref{fig_2} shows that law and order, $\mu(t)$, and $\nu(t)$ need over 50\% of the GDP just for keeping the economy in operational condition. The educational data of Japan were obtained from the reports of the Japanese Ministry of Education [7]. They include every contribution from kindergarten to private schools and universities. The other data are obtained from the national economic reports. The law and order data include every relevant expense from the legislative to the military. That they are close to and increase with the capital functions is mainly due to the  increasing inequality of the distribution of wealth. The plots for $\mu(t)$ agree automatically with their data since the latter follow from the data for PC and GDP whose agreement with theory is shown with figures \ref{fig_3} and \ref{fig_4}.

The following solutions apply to the industrial addition to the very small initial agricultural level $y_o$. Since most of the total cost and profit of $k_s=k-k_w$ appear also in its year of construction the national  growth dynamics must be derived with the total value $k$ of PC. Its annual change is from (\ref{eq_13}) for constant $G$ given by
\begin{equation}
 \label{eq_14}
\dot{k}=\mu G \dot{y}+\dot{\mu}Gy.
\end{equation}                                                 
With (\ref{eq_13}) and (\ref{eq_14}) the total support for PC consists of the sum
\begin{equation}
 \label{eq_15}
\dot{k}+k/G=\mu(1+G\dot{y}/y)y+\dot{\mu}Gy \equiv \overline{\mu}y+\dot{\mu}Gy.
\end{equation} 

The gross fixed capital share $\overline{\mu}$ corresponds to economics’ national saving constant for PC. It includes the maintenance of the accumulated PC and that part of its annual addition that comes naturally with the annual addition of the GDP. It does not include the term $\dot{\mu}Gy$ that allows increasing the gross capital coefficient $k/y=\mu G$ from its 18th century agricultural level of $\mu_o\overline{\varepsilon}G=1$ to China’s present top level of $\overline{\mu}G\overline{\varepsilon}=8$. 

$\dot{\mu}$ drives economic growth especially in the initial phase of industrialization when $\mu(t)$ is still small. Then PC and GDP are diverging. This is compensated later with a decreasing $\dot{\mu}$ until divergence and further growth stop at the asymptotically reached maximal state. Then $\overline{\mu}$ is entirely used up for maintaining the accumulated value $\overline{k}=\overline{\mu}G\overline{y}$. This includes new technology for replacing outdated PC. So far every diligent nation grew exactly in this way. It is the optimal growth path because it does not waste any resources by maintaining the original value of PC and recovering as fast as possible with the nation's maximum value of $\overline{\mu}$. Fiscal depreciation and reinvestment in required PC stabilize this path. 
 
The capitalization term $\dot{\mu}Gy$ is clearly seen in the national data for the fast recoveries of Japan and West Germany from World War II. Unfortunately it was neglected for the neoclassical growth theory of 1956, not least because it would have excluded the convenient steady state solution of exponential growth. Nevertheless,  (\ref{eq_15}) reduces to solving 
\begin{equation}
 \label{eq_16}
\mu\left(1+G\dot{y}/y\right)=\overline{\mu}.
\end{equation}                                                       
A constant $\mu$ yields Solow's steady state solution, specifying even its exponential growth rate $(\overline{\mu}/\mu-1)/G$ for $y$ and $k$. But the optimal growth path's use of $\overline{\mu} y$ in (15) for increasing the total capital coefficient $k/y=\mu G$ $and$ maintaining the accumulated value of $k$ leads to the inescapable trade-off (16) between the growth rate and $\mu$, $y$ in (25), and national wealth $k$ in (19) with (20).

The typical business cycle policy of stimulating growth by injecting short term money for PC is conterproductive for the optimal growth path. Not knowing the trade-off (16) lead after convergence within 28 years to the huge gap between public debt and private fortunes and to the banking crash of 2008. It must not be repeated since (19) predicts with (20) the required investment.  

The only analytic solution for this optimal growth path with two storing loops is the S-function 
\begin{equation}
 \label{eq_17}
y=\frac{\overline{a}}{1+e^{(T-t)/E}+e^{\beta(\tau-t)}}
\end{equation}                                           
where $\tau$ is the halftime and $\beta$ the initial growth rate of recovery. For times larger than $\tau$ (17) converges into the simple S-function 
\begin{equation}
 \label{eq_18}
a=\frac{\overline{a}}{1 + e^{(T-t)/E}}
\end{equation}                                                     
named industrial evolution. Its growth parameter $1/E$ picks up the upper limit of HC for the GDP (9) with (10). The halftime $T$ and the maximum GDP per capita $\overline{a}$ follow with the information obtained from  figure 4.

 S-functions are the irreversible analogue to periodic Sinus functions. Both disclose intrinsic time constants with their time and phase sifts, respectively.
The solution for $\mu$ and PC is given  with (13) by
\begin{equation}
 \label{eq_19}
k=\mu Gy=\overline{\mu}Gy_k
\end{equation}                                                   
and
\begin{equation}
 \label{eq_20}
y_k=\frac{\overline{a}}{1+e^{T+\Delta T-t}+e^{\beta(\tau+\Delta\tau-t)}}.
\end{equation}  
The delay of PC with respect to the GDP is due to the decay of the accumulated PC while new PC is built up. The delay times are obtained by inserting (17) and its time derivative into (16) with the results 
\begin{equation}
 \label{eq_21}
\Delta T=E \log \left(1+G/E\right)=21\; \text{ years}
\end{equation}                           
and
\begin{equation}
 \label{eq_22}
\Delta \tau=\beta^{-1} \log \left(1+\beta G\right).
\end{equation}  
                                      
The growth parameter of recovery from war or catching up with the G7 from the agricultural level is according to the initial state of (16) given by
\begin{equation}
 \label{eq_23}
\beta=\left(\overline\mu/\mu_e-1\right)/G.
\end{equation}         
The initial value $\mu_e$ of the capital function follows from (\ref{eq_23}) since $\overline{\mu}$ is nationally fixed and $\beta$ is directly observed during the initial exponential phase of the S-function (\ref{eq_19}). $\mu_e Gy$ is so to speak the “entrance fee” for catching up with the G7 because starting far below the technical state of the art loses time and resources. Resolving (23) for $\overline{\mu}$ yields the required support for recovering with a desired initial growth rate $\beta$.

Figure 3 shows three examples for the measurement of $E$ and $G$ with the time shifts (21) and (22). Japan’s well-known overvaluation of real estate is quantitatively disclosed by the theory. It is corrected by using the measured time shift (22) and the actual growth parameter $\beta$ with West Germany's value for $\mu_e$. The national halftimes for recovery, convergence parameters, and the initial and final values of the capital function are listed in table \ref{tab_1}. 

\begin{table}[b]
\centering\begin{tabular}{|l|c|c|c|c|}
\hline
Country       &   $\tau$       &    $\overline{\mu}$   &  $\mu_e$        &     $\beta$ \\ 
\hline
USA       &      1965    &    0.18   &     0.08     &     0.05 \\ 
\hline
Germany   &	 1970    &    0.25   &     0.08     &     0.09 \\
\hline
Japan     &      1971    &    0.26   &     0.08     &     0.09 \\
\hline
Korea     &      2010    &    0.27   &     0.09     &     0.08 \\
\hline
China     &      2040    &    0.34   &     0.11     &     0.10 \\
\hline
\end{tabular}
\caption{\label{tab_1} The national parameters for optimal recoveries}
\end{table}

(17) and (18) are plotted in figure 4 with their final amplitude of $\overline{a}=75.000$ US\$ in the value of 1991, the halftimes $\tau$, and the industrial evolution’s halftime T=2040.  The data are five-year averages of the annual data taken from the national statistical reports.

The industrial evolution (18) was obviously established in 18th century UK as the first industrial nation. Its 40 year stagnation around 1900 is also seen in the working time of figure 1. It was due to empire problems and increasing competition from the Continent and the USA. (18) is since the Great Depression the goal of convergence for all recoveries. Its immunity to all disasters of the 20th century is quantitatively explained when the indestructible human capacity has the reaction time $E$ to changing existential conditions. The few nations with GDPs above (18) owe their relative position to a preference for working with non-physical capital, i. e., with financial transactions, but their wealth and their stablity depend on the success of structurally balanced economies. 

The physical theory's solutions (17) and (18) are linked with 

\begin{equation}
 \label{eq_24}
y^{-1}=a^{-1}+\overline{a}^{-1}e^{\beta(\tau-t)}.
\end{equation} 
The third term is the decreasing difference between the inverted S-functions. This means there exists a simple exponential decay of "human incapacity" $1/a(t)$ to a lowest industrial level  $1/\overline{a}$. (24), (17), and (18) are the integrals of the growth laws for the national growth rate
 
\begin{equation}
\label{eq_25}
\dot{y}/y=\beta(1-y/a)+(\dot{a}/a)y/a
\end{equation} 
and the industrial evolution's growth rate  

\begin{equation}
 \label{eq_26}
\dot{a}/a=(1-a/\overline{a})/E.
\end{equation}  
(25) allows determining the convergence parameter $\beta$ of a nation from any point $y(t)$ with slope $\dot{y}$ and associated value of $a(t)$. (25) and (26) show directly the optimal growth path's S-functional trade-off between growth rate and GDP. 

This theory has no adjustable parameters. The national parameters for the support of physical capital and human capacity are known from the national reports. $E$ and $G$ are directly measured systemic constants. The halftimes of the industrial evolution and national recoveries are set by history. They are easily measured with the precision of one year in the linear phase of every S-function. The amplitude $\overline{a}$ of the industrial evolution seems to be the only exception because at this point it is extrapolated. But in due time we found with the unisex life expectancy $L(t)$ of the pioneering nations a biologic process with the same growth parameter [8].

$L(t)$ is like longevity to a relevant extent heritable [9]. The medical, natural, and technical sciences were and will always be closely linked. Nothing works against the laws of nature. There is no intrinsic contradiction between $h_s$, $L(t)$, and $a(t)$, but there is a causality problem between $L(t)$ and $a(t)$ unless HC controls both cultural quantities simultaneously.    

The data for $L(t)$ are included in figure 4. The lower initial values for Germany and Japan are due to higher child mortality because of their later entry into the industrial society. The plot in figure 4 is the simple S-function
\begin{equation}
\label{eq_27}
L=L_o +\frac{(\overline{L}-L_o)}{1+e^{(T_L-t)/E}}.
\end{equation} 
 The halftime is $T_L=1981$. Since the industrial evolution's halftime is $T=2040$ the presently active G7 generation lives in the middle between the maximum annual increases of both quantities. In figure 1 they join annual working time as cultural variables of the industrial society. $L_o=30$ years was since antiquity the reproduction minimum for stabilizing the population in times of peace. The necessarily shorter reproduction cycle must be nearly identical with the physical lifetime $G=25$ years of PC. Industrial generations mainain the value of their inherited PC obviously by replacing it on average just once in their lifetime.

$L(t)$ and $a(t)$ are comparable because life insurers correct the mortality tables for all deaths due to identifyable disasters and $a(t)$ provides the existential conditions above all disasters. 
$\overline{L}=118$ years specifies the maximum unisex life expectancy believed since antiquity to be about 120 years. It is not just extrapolated, it is measured with the precedence of $L(t)$ with respect to $a(t)$ of $T_L-T=\overline{L}/2=59$ years, the length of the line between both inflection points in figure 4. This precedence was shown to follow quantitatively without any adjustable parameter when human life integrates and averages over the existential conditions given by the industrial evolution $a(t)$ [8].

This means human life solves not only economics' absolute problem of measuring and integrating the benefits of using the produced goods and services of the GDP, it also records the result per capita with a unit that is independent of inflation. The ratio $(L-L_o)/a$ is constant for the full range of both variables so that $\overline{L}$ and $\overline{a}$ are constants for at least the industrial society.

Since the educated knowledge  cannot be duplicated just with money the economic value of HC can be derived directly from (12) and the analogy 
\begin{equation}
 \label{eq_28}
\nu=\overline{\nu}/(1 + E \dot{a}/a)
\end{equation}                                                       
to (16). The “Sputnik Shock” increase of $\nu(t)$ in figure 2 between 1958 and 1975 created the largest technology push of all times. It is known as the Route 128 or the Silicon Valley phenomenon for its largest academic and industrial research parks around Boston and the San Francisco Bay. But its economic effect was negligible, the G7 growth rates continued their saturating trend. 

For the initial state $\dot{a}/a=1/E$ and $\nu_o$ from (\ref{eq_6}) one obtains from (28) for the gross fixed capacity share $\overline{\nu}=2\nu_o=2/\overline{\varepsilon}E$. The capacity function in figure \ref{fig_2} is plotted with $\overline{\nu}=4\nu_o=4/\overline{\varepsilon}E$ because that is closer to the data without the Sputnik Shock. This yields $2y_o/\overline{\varepsilon}<h(t)<4a/\overline{\varepsilon}$. According to (5) the actually required part of HC is $h_s=y/(\overline{\varepsilon}-w)$. With $w=y/k_w=1/\mu_wG$  one gets
\begin{equation}
 \label{eq_29}
 h_s=y/\overline{\varepsilon}(1-1/\mu_w\overline{\varepsilon}G)<1.4a/\overline{\varepsilon}\approx(1/3)h(t)
\end{equation}    
when Germany and the UK are used with their $\mu_w=0.15$ as numerical example.

This is the first quantitative assessment of the economic effectiveness of G7 educational systems. A missing factor of 3 would finish this last part of the theory because the educated value of HC could not achieve the G7 level. A safety factor of 3 and all man made disasters suggest that the immunity of the industrial evolution over centuries is not due to education. It could be credited to the biologic adaptability of HC, i. e., to a genetically stabilized property.

\section{Application and first consequences} 

This theory will be used for quantifying and predicting the medium and long-term growth of every industrializing nation. Inserting the 5 year averages of a nation’s real GDP per capita into figure 4 averages out short term fluctuations, identifies the nation’s position $y(t)$ and the corresponding value of $a(t)$, and shows how well the nation is tuned to the optimum path. This yields with (25) the growth parameter $\beta$ so that the entire GDP (17) can be plotted with an approximate value for $\tau$. Correcting for the observed time shift yields the nation’s actual halftime. The last step is comparing the nation’s physical capital data with the plot of (19) computed with the nation’s official value for $\overline{\mu}$ and (19) to (22). With $\beta$, $\mu_e\approx0.10$, and (23) the lower time shift (22) should  confirm the nation's reported gross fixed capital share $\overline{\mu}$. 

The US\$ value of figures 3 and 4 and the exchange rates of 1991 are due to the time of reference used by the statistical reports after convergence of the G7 with the industrial evolution. The same procedure can of course be followed with the nation’s inflation and population corrected values and the current value and exchange rate for $\overline{a}$. 

A nation not satisfied with its course must renormalize its economy, but only two national parameters are in the end responsible and accessible.  The total support $\overline{\nu}$ for the indestructible human capacity with 5 to 6\% of the GDP is required for reaching and staying at the level of the industrial evolution. The total support $\overline{\mu}$ for the destructible physical capital sets the pace for recovery. Successful nations set their largest possible $\overline{\mu}$ and keep it constant. Fast change causes inefficient overshoot and the result of slow change is difficult to observe in one or two legislative periods. 

Behind all facts and physical science the success of the G7 was implicitly achieved by creating the liberal conditions allowing human capacity to organize and amplify spare time. The explicit policy supported mainly the accumulation of physical capital with hard work. The physical theory adds now the insight that fighting human capacity's superior stabilizing and limiting role is counterproductive. 

Germany and Japan had the educational level required for fast recovery from World War II, and Korea had it after the Korean War. But after the Long March and the Cultural Revolution the PR of China could not have the required educational level. Initially its fast growth was due to its very tough joint venture policy with G7 companies. It began when the Cold War was not yet over.  The USA supported an independent China. Germany and  Japan needed a new big market. After 1980 they observed but misunderstood their sudden "weakness of growth" called "hollowing out" in Japan. The physical theory shows that both were caused by the sharp drop of the demand for investment capital in the transition from PC to HC, i. e., from large recovery growth rates into the small growth rate of the industrial evolution named convergence crisis. 

China made perfect use of this unique window of time. With joint ventures across the global  business world and returning Chinese nationals it absorbed the G7 policy under a dramatically changing political ideology. Its presently observed transition from exponential to linear recovery and this century's largest imaginable convergence crisis between 2040 and 2050 were predicted and communicated as inevitable consequences of fast recovery under the control of essentially the same human capacity.

Figure 4 shows the present position of the G7 type nations between the inflection points of the life expectancy and the industrial evolution. This is a factor of 30 above 18th century agricultural GDPs and a factor of 2.5 below the asymptotically approached maximum GDP. For the G7 the economic problem is solved, as predicted for the ensemble average [6]. But the industrial society's social problems are far from being solved. They accumulated in the G7 after convergence while waiting for higher growth rates. They should have been and must now be solved at any growth rate. 

The industrial society's basic differential equation (16) and its solution (18) quantify the inescapable trade-off between growth rate and wealth. China has still one generation gap with comfortable growth rates ahead. When its demand for energy follows G7 history China will more than double the global annual consumption of fossil reserves until 2040. Replacement technologies will increasingly challenge the global research community. As the only alternative it is good to know that S-functions can also control reducing untenable levels of physical capital, especially for nations with decreasing population.

For two generations the G7 enjoyed peace on their soil. They were not limited by labor, global resources, technical progress, or investment money. But even their largest growth experiments could not stop the decrease of their growth rate from an average 5 \% p.a. in 1950 to below zero after 2008. The largest technology push of all times from 1958 to 1975 was quantitatively described with figure 2. The largest performance push of all times from 1975 to 2008 put every business activity under the dictate of short term monetary success.

The long-term consequences of the monetary growth fetishism are still too opaque for prediction. But the total failure of the economically strongest nations to at least stop the saturation of their growth rates by capitalism's toughest experiments, and the cultural variable’s immunity even to world wars, suggest that the process stabilizing  and limiting the industrial evolution may be credited to the inherited component of human capacity. Specifically it may just be due to mankind's basic difference in the origin of species: the limited adaptability to the changes of the existential conditions mankind's unique brain can achieve. Otherwise the constancy of $E$, of the related $G$ and $L_o$, and of the related $\overline{L}$ and $\overline{a}$ over centuries and orders of magnitude in the GDP per capita would indeed require Adam Smith's invisible hand. 

The constituting property of the simple S-function (26) is the proportionality of its growth rate to the remaining part of its entire path. The proportionality constant is its growth parameter. It is the constant rate of decay of the inverse S-function to a minimum level of "incapacity". A zero level would mean human omnipotence and unlimited exponential growth. But a system knowing according to (26) by its current growth rate continuously where it stands with respect to its initial and final state may be a priori programmed like reproducing life. This simple fact allowed life insurers predicting the life expectancy after birth better than any other long-term process could be predicted. So far they could only assume that both the biologic and the economic trend are stable and closely related. Both processes have not only the same constituting property but also the same growth parameter and the theoretically expected precedence of the life expectancy. 

Although the physical theory describes and predicts the peaceful economic growth without adjustable parameter the future of the industrial society is not as deterministic as individual life. The theory shows how mankind can use its long-term choice between producing more affluence with work and physical capital or pursuing happiness with spare time and human capacity.

\end{document}